\documentclass[12pt]{article}
\usepackage{times}
\usepackage{geometry}
\usepackage[utf8]{inputenc}
\usepackage{enumitem,amssymb, color}
\usepackage{ragged2e}
\newlist{thematic}{itemize}{8}
\setlist[thematic]{label=$\square$}
\usepackage{pifont}

\usepackage{graphics,graphicx}
\usepackage{sidecap}

\begin{document}
\raggedright
\huge
Astro2020 Science White Paper \linebreak

Resolving the cosmic X-ray background with a next-generation high-energy X-ray observatory   \linebreak
\normalsize

\noindent \textbf{Thematic Areas:} \hspace*{60pt} $\square$ Planetary Systems \hspace*{10pt} $\square$ Star and Planet Formation \hspace*{20pt}\linebreak
$\boxtimes$ Formation and Evolution of Compact Objects \hspace*{31pt} $\boxtimes$ Cosmology and Fundamental Physics \linebreak
  $\square$  Stars and Stellar Evolution \hspace*{1pt} $\square$ Resolved Stellar Populations and their Environments \hspace*{40pt} \linebreak
  $\boxtimes$    Galaxy Evolution   \hspace*{45pt} $\square$             Multi-Messenger Astronomy and Astrophysics \hspace*{65pt} \linebreak
  
\textbf{Principal Authors:}

Name: Ryan C.\ Hickox
 \linebreak						
Institution: Dartmouth College
 \linebreak
Email: ryan.c.hickox@dartmouth.edu
 \linebreak
Phone: (603) 646-2962 
 \linebreak
 
 Name: Francesca Civano	
 \linebreak						
Institution: Center for Astrophysics $\vert$ Harvard \& Smithsonian
 \linebreak
Email: fcivano@cfa.harvard.edu
 \linebreak

\textbf{Co-authors:} D. Ballantyne (Georgia Tech), M. Balokovi\'{c} (Center for Astrophysics $\vert$ Harvard \& Smithsonian), P. Boorman (University of Southampton), W. Brandt (Pennsylvania State University), R. Canning (Stanford University), F. Fornasini (Center for Astrophysics $\vert$ Harvard \& Smithsonian), P. Gandhi (University of Southampton), M. Jones (Center for Astrophysics $\vert$ Harvard \& Smithsonian), G. Lansbury (University of Cambridge), L. Lanz (Dartmouth College), G. Lanzuisi (INAF, Bologna, Italy), K. Madsen (Caltech), S. Marchesi (Clemson University), A. Masini (Dartmouth College), T. Ananna (Yale University), D. Stern (JPL, Caltech), C. Ricci (UDP, Chile)
  \linebreak

\textbf{Abstract:}
The cosmic X-ray background (CXB), which peaks at an energy of $\approx$~30 keV, is produced primarily by emission from accreting supermassive black holes (SMBHs). The CXB therefore serves as a constraint on the integrated SMBH growth in the Universe and the accretion physics and obscuration in active galactic nuclei (AGNs). This paper gives an overview of recent progress in understanding the high-energy ($>$~10 keV) X-ray emission from AGNs and the synthesis of the CXB, with an emphasis on results from NASA's {\em NuSTAR} hard X-ray mission. We then discuss remaining challenges and open questions regarding the nature of AGN obscuration and AGN physics. Finally, we highlight the exciting opportunities for a next-generation, high-resolution hard X-ray mission to achieve the long-standing goal of resolving and characterizing the vast majority of the accreting SMBHs that produce the CXB.

\pagebreak

\justify

\section{Introduction}

The cosmic X-ray background (CXB) was first discovered by the earliest X-ray astronomical rocket flights (Giacconi et al. 1962), and over the past 50 years, its origin has since been a major research area in high-energy astrophysics. The CXB is now known to be primarily composed of emission from individual active galactic nuclei (AGNs; e.g., Hickox \& Alexander 2018), whose X-ray emission provides key constraints on the cosmic evolution of the supermassive black holes (SMBHs). These SMBHs play an important role in the evolution of galaxies and large-scale structure (see Civano et al.\ 2019 White Paper). The CXB spectrum peaks at an energy of $\approx$30 keV, indicating a significant contribution from heavily obscured AGN whose emission is attenuated at lower energies and show strong signatures of Compton reflection. In recent years, great progress has been made (particularly with the {\em Chandra}, {\em XMM-Newton}, {\em Swift}, and {\em NuSTAR} observatories) in understanding the high-energy emission from AGN with subsequent insights into the process of SMBH accretion, the impact of SMBHs on galaxy evolution, and the ultimate composition of the CXB. However, the challenges of observations at hard ($>$~10 keV) X-ray energies mean that {\em direct} knowledge of the AGN that produce the bulk of the CXB has remained elusive. In this White Paper, we present an overview of our current understanding of the origin of the CXB and the nature of hard X-ray emission from AGN, with a focus on results from {\em NuSTAR} (Harrison et al.\ 2013). We next discuss current challenges and open questions regarding the nature of AGN obscuration and AGN physics. Finally, we present some of the exciting opportunities for progress with a future high-resolution hard X-ray mission.

\section{Black hole evolution and the origin of the CXB}

Prior to the launch of {\em NuSTAR}, our understanding of the $>$10 keV CXB came almost entirely from wide-field observatories with limited resolving power (see e.g., Gilli, Comastri \& Hasinger 2007; Ajello et al.\ 2008). In soft ($<10$ keV) X-rays, sensitive, high-resolution observatories, specifically {\em Chandra} and {\em XMM}, provided detailed measurements of the individual AGN that made up the CXB, resolving up to 80\% of the $<$2 keV CXB in the deepest observations (Hickox \& Markevitch 2006; Xue et al.\ 2012). {\em Chandra} and {\em XMM} surveys and associated multiwavelength observations yielded constraints on the evolution of both the number counts of AGN (e.g., Luo et al.\ 2017) and of their X-ray luminosity function (XLF; e.g., Aird et al.\ 2015a). Together, these observations have informed {\em synthesis models} of the CXB, in which an evolving population of AGN is responsible for the integrated CXB spectrum (e.g., Gilli, Comastri \& Hasinger 2007; Ueda et al.\ 2014; Aird et al. 2015b; 
Jones et al. 2017; Ananna et al.\ 2019; Figure 1). CXB synthesis models have led to a number of important conclusions about the AGN population, most strikingly that the total CXB spectrum is peaked toward higher energies than most of the individually detected AGN, indicating the presence of a large ``hidden'' population of heavily obscured (Compton-thick, $N_{\rm H}>10^{24}$ cm$^{-2}$), X-ray hard AGN (e.g., Gilli, Comastri \& Hasinger 2007).

The major challenge in interpreting CXB synthesis models has come from the limited observational constraints on the hard ($>10$ keV) X-ray emission for individual AGN. Until 2012, some of the most sensitive observations came from the {\em Swift} Burst Alert Telescope (BAT) and {\em INTEGRAL} observatories,which are able to obtain detailed spectroscopy at energies between 14 keV and  as high 195 keV and so provided us a valuable picture of the high energy emission for bright, nearby AGN. However, {\em Swift}/BAT's and {\em INTEGRAL}'s limited sensitivity means that they can only detect the absolute most luminous sources beyond the local Universe ($z\sim $0.6). In 2012, the launch of {\em NuSTAR} (Harrison et al. 2013), the first focusing X-ray observatory at energies $>10$ keV, enabled a $>$100-fold increase in sensitivity, allowing us to probe a more complete population of AGN as well as its evolution with redshift.

To capture this AGN population, {\em NuSTAR} has carried out an extragalactic survey program consisting of several blank fields of varying depths and areas that cover well-studied multiwavelength survey regions (e.g., COSMOS, Extended Chandra Deep Field South, Extended Groth Strip, UKIDSS Ultra Deep Survey, Hubble Deep Field North). The widest area component is the Serendipitous Survey, which searches for sources in the fields of pointed {\em NuSTAR} observations and involves dedicated spectroscopic follow-up. Together, these surveys have yielded $\sim$~1000 hard X-ray selected AGN and have led to the most precise measurements to date of the hard X-ray flux distribution ($\log N$--$\log S$; Harrison et al.\ 2016) and X-ray luminosity function (Aird et al.\ 2015a). An ongoing {\em NuSTAR} survey program extends this analysis by targeting known X-ray bright sources in wider soft X-ray fields (e.g., XBo\"{o}tes and Stripe82X).

\begin{figure}[t]
\begin{center}
\includegraphics[width=0.48\textwidth]{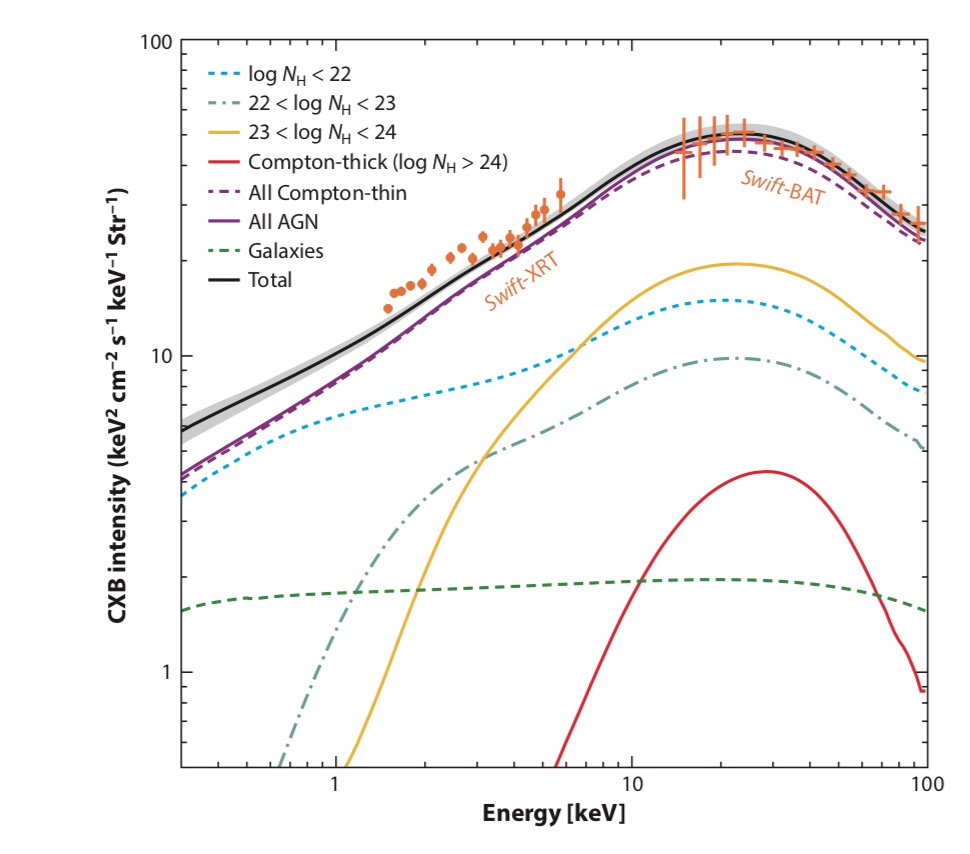}
\includegraphics[width=0.48\textwidth]{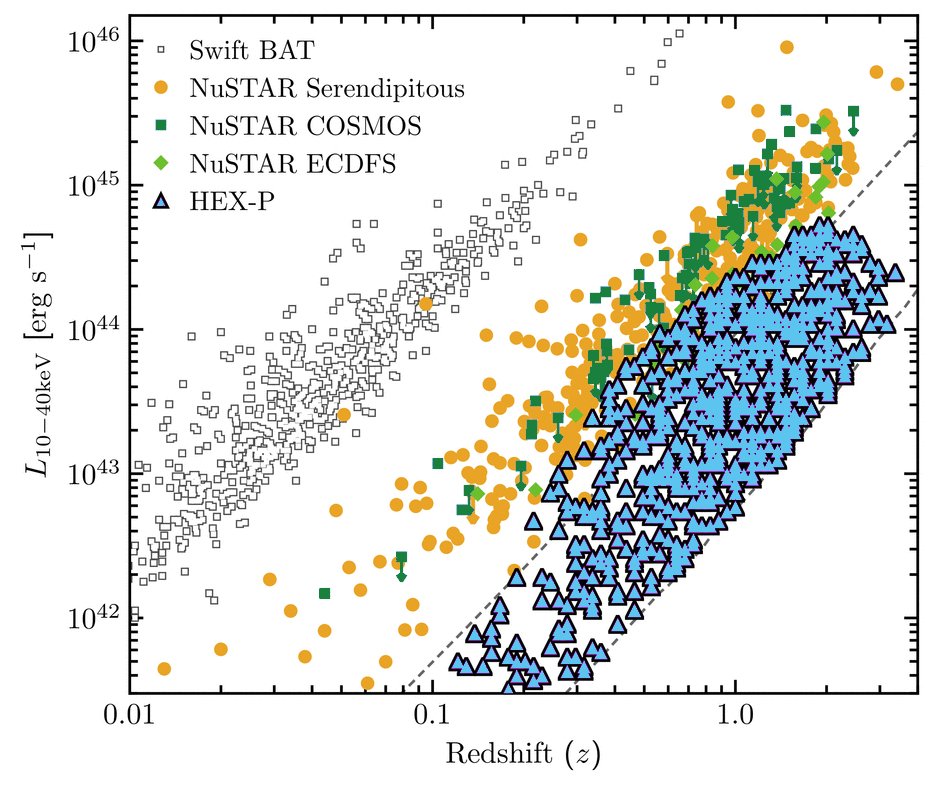}
\end{center}
\vspace{-1.0cm}\caption{\footnotesize{{\em Left:}  CXB synthesis model (Aird et al. 2015b) showing the contributions from unobscured AGN (blue dotted line) and obscured AGN with varying levels of N$_{\rm H}$. Obscured sources dominate the high-energy peak of the CXB , with a significant contribution from Compton-thick AGN (thick red line). {\em Right:} Distributions in redshift and X-ray luminosity for current hard X-ray surveys by {\em Swift/BAT} and {\em NuSTAR} and the range that could be probed by the planned X-ray telescope {\em HEX-P} assuming a similar shallow-medium-deep survey strategy.}}
\label{lxz}
\end{figure}

{\em NuSTAR} observations have also dramatically advanced our understanding of the connections between obscuration, accretion physics, and the high-energy X-ray emission from AGNs. A targeted survey of nearby obscured systems has led to key constraints on the nature of X-ray reprocessing from the obscuring ``torus'' (e.g., Gandhi et al.\ 2014; Marchesi et al. 2018, 2019; Balokovi\'c et al. 2018; Lanz et al. 2019), indicating a potentially wide range of covering factors among obscured AGN at a given luminosity. Similar observations have uncovered AGN that appear to be intrinsically X-ray weak (Luo et al.\ 2014) as well as sources with very complex, multi-phase obscurers (Bauer et al. 2015, Teng et al. 2015) and very heavy (Compton-thick) obscuration (e.g., Koss et al.\ 2016) that may preferentially be associated with late-stage galaxy mergers (e.g., Ricci et al.\ 2017a; Lansbury et al.\ 2017).

Finally, {\em NuSTAR} has observed higher-redshift ($z>0.1$) luminous, obscured AGN that are initially identified at other wavelengths, for example through high-excitation optical lines (Lansbury et al. 2014, 2015) and {\em WISE} mid-infrared colors (Stern et al.\ 2014, Yan et al.\ 2019). {\em NuSTAR} has shown that many previously X-ray detected AGN are significantly more obscured than can be determined through soft X-rays alone; including constraints from {\em NuSTAR} indicates a large Compton-thick fraction of $\sim$30\% or higher (Lansbury et al.\ 2014, 2015). Furthermore, some luminous obscured AGN with {\em no} previous X-ray detections are extremely weak or undetected in deep {\em NuSTAR} exposures, implying $N_{\rm H}$ of $10^{25}$ cm$^{-2}$ or greater (Stern et al. 2014, Yan et al.\ 2019).

These results all inform the inputs to CXB synthesis models, by providing information on the X-ray spectral shapes, $N_{\rm H}$ distribution, and luminosity evolution of AGN. Recently, we have been able to perform  direct tests of CXB synthesis models, by stacking the {\em NuSTAR} emission from known AGN (Hickox et al.\ in prep; Figure 2). These results are able to distinguish between different CXB synthesis models and favor the latest prescriptions that include sophisticated handling of the distributions and evolution of spectral parameters (e.g., Ananna et al.\ 2019).

Looking toward the future, we note that {\em NuSTAR} surveys have dramatically increased the fraction of the hard CXB that is resolved into individual sources, to $\approx$30\% (Harrison et al.\ 2017) compared to $<1\%$ with {\em Swift}/BAT. However, {\em NuSTAR} detections are still mainly limited to $<16$ keV due to instrumental backgrounds, and {\em NuSTAR} is approaching a fundamental flux limit for direct detection due to source confusion. Thus the majority of the CXB is still not resolved directly, providing a large discovery space for more sensitive hard X-ray observations. Below, we outline some outstanding challenges in our understanding of the hard X-ray emission from AGN and opportunities with a future sensitive, high-resolution hard X-ray mission.

\section{Current challenges and open questions: AGN obscuration, connections to galaxies, and accretion physics}

A major challenge in AGN population studies with {\it Chandra} and {\it XMM-Newton} is the impact of absorption, which preferentially affects soft (0.5-2 keV) X-rays and can introduce significant uncertainty on obscuration levels and intrinsic luminosities without sensitive high-energy constraints. 

Thanks to the large sample provided by {\em NuSTAR} surveys (Figure~1), it was possible to measure for the first time the hard X-ray luminosity function (Aird et al. 2015a) beyond the local Universe and the evolution of the obscuration distribution (Zappacosta et al. 2018, Del Moro et al. 2017) to $z=$3. Nonetheless, the sample is limited in both size and depth. Moreover, most of the sources are not detected in the hardest 8--24 keV band, and their luminosities are all probing the brighter range above the ``knee'' of the luminosity function, missing the bulk of the population contributing to the total accretion density. While {\it NuSTAR} is continuing to perform surveys of several fields and the serendipitous sample is increasing continuously (now $\sim$ 1000 sources), improving the depth is a key component. Measuring sources below the ``knee'' of the luminosity function at $z>$0.5 and probing for the first time the full range of luminosity and $N_{\rm H}$ at all redshifts would allow us to constrain the total accretion density at the energy peak of the CXB and beyond. 

A related question is the physical nature of the obscuration, arising from a parsec-scale ``torus'' around the AGN central engine (e.g., Netzer 2015; Ricci et al.\ 2017b), or from a range of scales from broad-line region clouds to galaxy-scale gas and dust lanes on the scale of the galaxy (e.g., Buchner et al.\ 2017; Hickox \& Alexander 2018; Blecha et al.\ 2018). Soft X-ray observations are often degenerate between obscuration by small-scale clumpy material and smooth, large-scale clouds, but hard X-ray observations can constrain the reflection component and provide important additional constraints on the geometry as well as the metallicity of the material, as the shape of the Compton hump depends on the abundances (e.g., Wilman \& Fabian 1999). 

An additional open question regards the presence of {\em extreme} Compton-thick obscuration ($N_{\rm H}>10^{25}$ cm$^{-2}$). While such sources are known locally (e.g., the canonical Seyfert 2 NGC 1068), their general abundance is poorly understood, in part because their X-ray emission, along with many other AGN signatures, is heavily suppressed. The existence of extremely obscured sources can strongly impact estimates of the total AGN power and the global radiative efficiency (e.g., Comastri et al.\ 2015). With sensitive hard X-ray observations of AGN selected in the optical, infrared, or radio, we can infer the presence of these heavily obscured sources (e.g., Yan et al.\ 2019; Aalto et al.
\ 2019 White Paper). To this end, {\em NuSTAR} is carrying out a dedicated survey of heavily obscured candidate AGN in the local Universe ({\em NuLANDS}; Boorman et al.\ in prep).

\begin{figure}
\begin{center}
\includegraphics[width=0.78\textwidth]{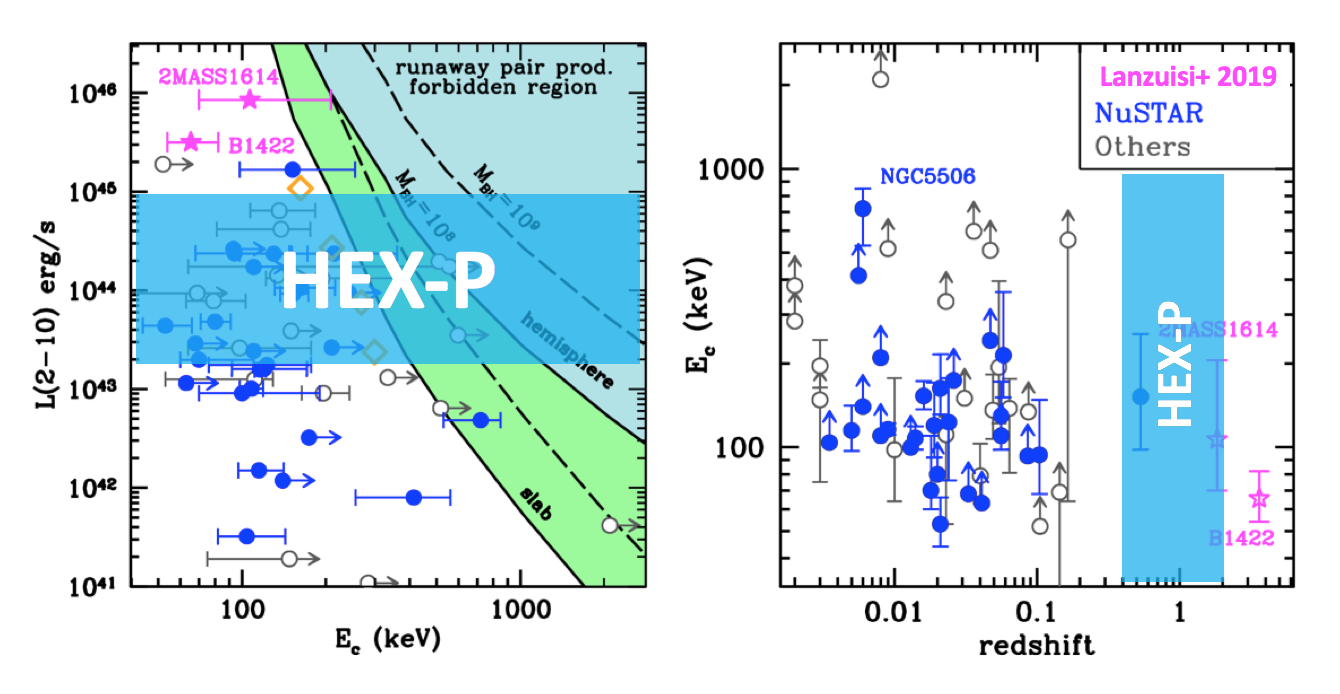}
\end{center}
\vspace{-1.0cm}
\caption{\footnotesize{High energy cut-off versus $L_X$ and $z$ for a sample of AGN from Fabian et al. (2015),  Lanzuisi et al. (2019), and Ricci et al. (2018). {\it NuSTAR} measurements are in blue and non-focusing hard X-ray telescopes are in gray. The region that {\em HEX-P} surveys could probe is highlighted in blue. Figure is adapted from Lanzuisi et al. (2019).}}
\label{coronaT}
\end{figure}

The characterization of the X-ray emission in AGN is the best tool available to investigate the physical properties of the innermost regions around accreting SMBH and to measure coronal properties (temperature, optical depth and geometry; see associated White Paper by Kamraj et al. 2019). X-ray reverberation mapping studies with {\it NuSTAR} have provided measurements for the corona radius, while measuring the cut off energy in X-ray spectra can provide the temperature (e.g., Ricci et al.\ 2018). Because of its limited bandwidth, {\em NuSTAR} cannot tightly constrain cut-off values larger than few hundreds of keV in the observed frame, except for sources with large photon statistics (see e.g. NGC 5506, Matt et al. 2015), and therefore these studies were so far limited to a sample of few tens of (mostly unobscured) sources (e.g., Fabian et al. 2015) up to $z<$0.1, luminosities $<$ 10$^{45}$ erg s$^{-1}$, and cut-off energies $<$200 keV. 
Only recently, the high-energy cut-off was measured for a few bright quasars with luminosities of 10$^{46}$ erg s$^{-1}$ in the 2--10 keV band at $z>$1. These measurements provided an estimate of the temperature, testing runaway pair production models by measuring coronal properties in high redshift sources where higher cut-off energies can be constrained as the spectrum is redshifted to lower energies (Lanzuisi et al. 2019; see magenta points in Fig. \ref{coronaT}). 

\section{Opportunities with a next-generation hard X-ray mission}

\begin{figure}
\begin{center}
\includegraphics[width=0.55\textwidth]{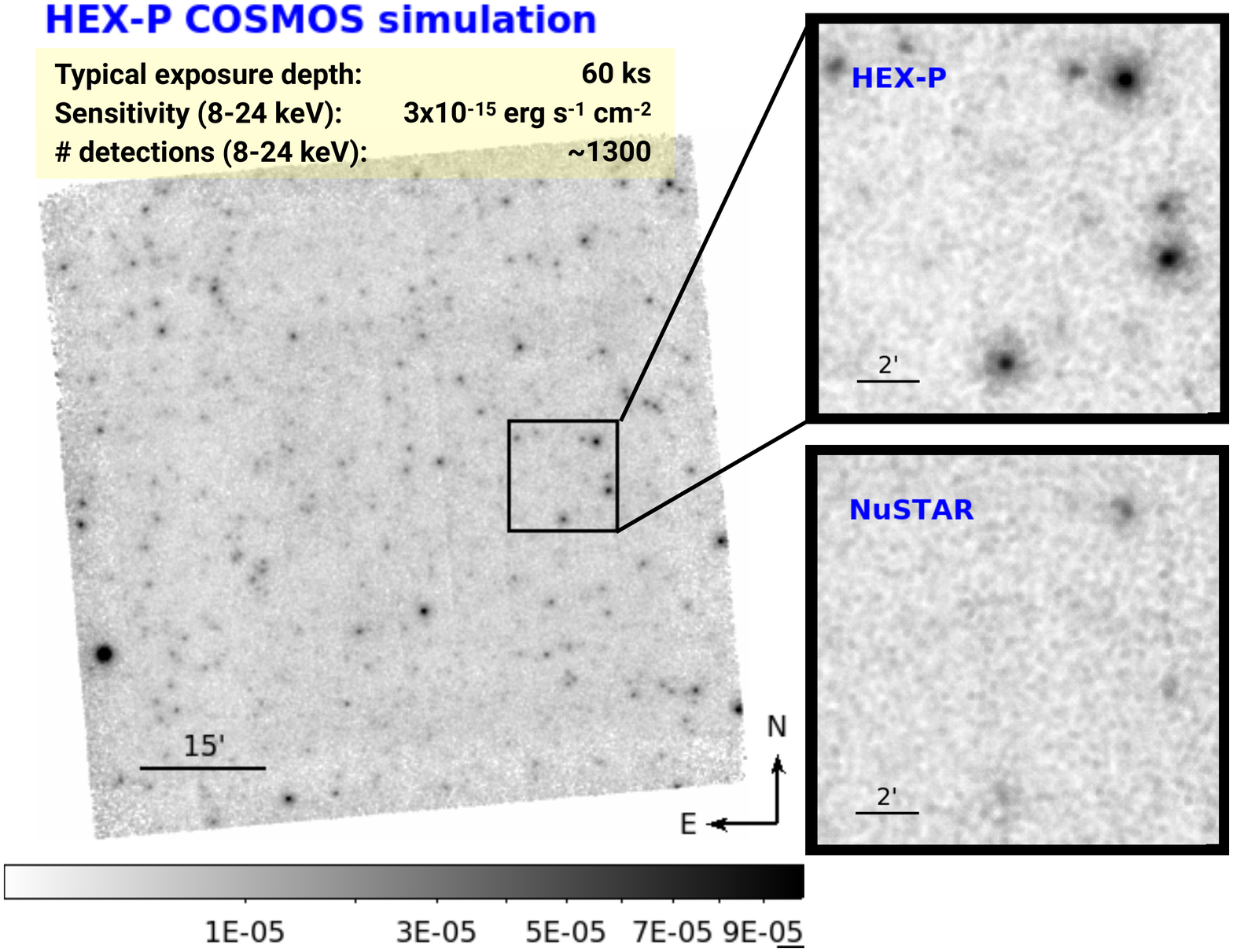}
\includegraphics[width=0.42\textwidth]{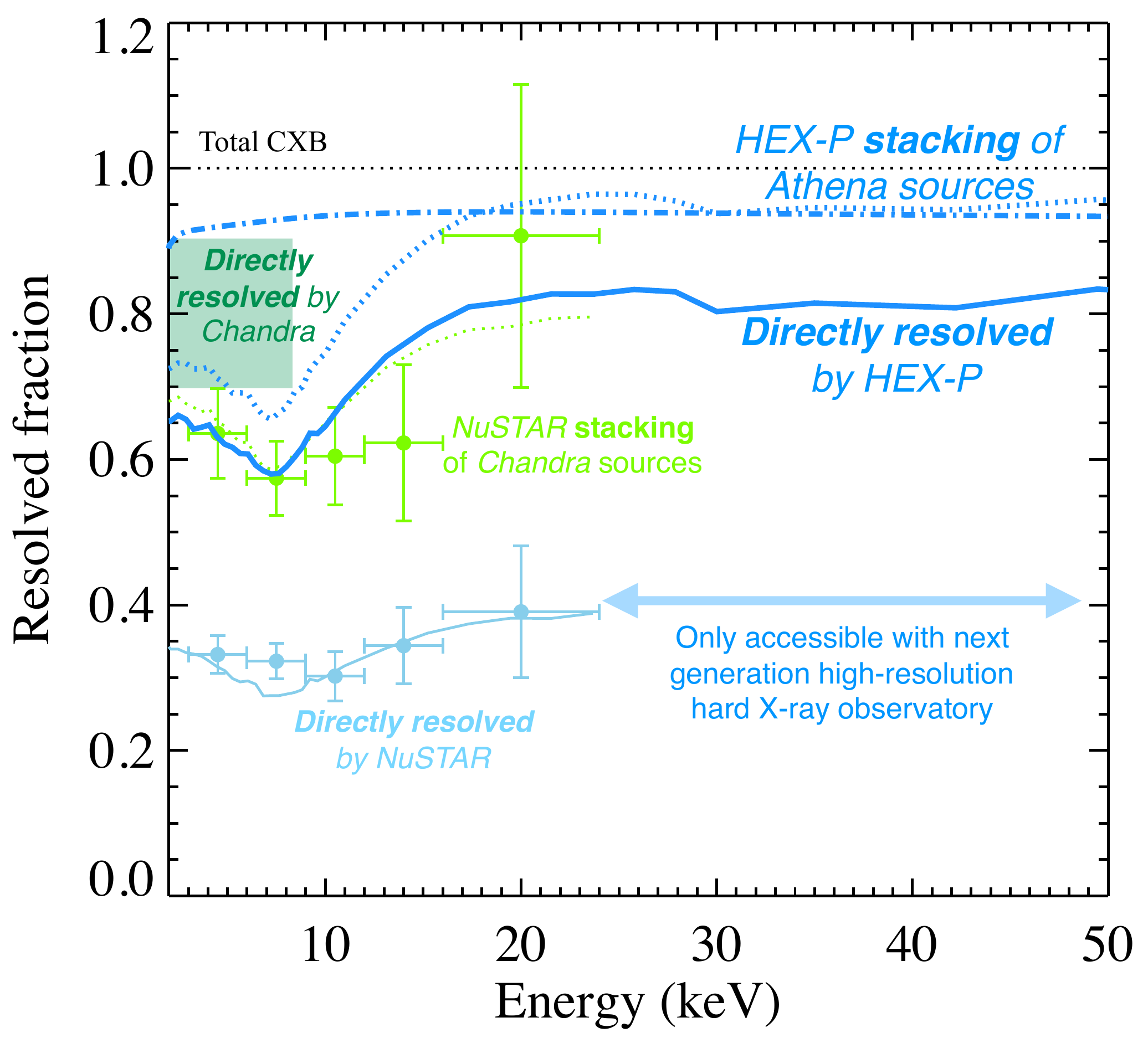}
\end{center}
\vspace{-1.0cm}
\caption{\footnotesize{{\em Left:} Simulation of a $\approx$2 deg$^2$ {\em HEX-P} field. {\em Right:} Fraction of the CXB vs.\ energy, resolved directly with {\em NuSTAR} (blue points) and {\em Chandra} (green box), and through {\em NuSTAR} stacking of {\em Chandra} sources (green points). Thick blue lines show predictions for the flux limits of {\em HEX-P}, or through {\em HEX-P} stacking of soft X-ray sources at the {\em Athena} limits. Models are from (solid and dotted) Ananna et al.\ (2019) and (dot-dashed) Jones et al.\ (in prep). } {\em HEX-P} can directly resolve up to 80\% of the hard CXB, and account for over 90\% through stacking of soft X-ray sources.}
\label{simul_image}
\end{figure}

The high scientific return of  {\it NuSTAR} and the opportunities described above for sensitive hard X-ray observations demonstrate the motivation for a High Energy X-ray Probe ({\em HEX-P}; Madsen et al.\ 2018), which would be  highly complementary to upcoming lower energy X-ray imaging and spectroscopy missions such as {\em XRISM} and {\it Athena} and concepts like {\it Lynx} and {\it AXIS}. The key requirements would be a wider energy bandpass (1 to 200 keV) and higher sensitivity at harder energies, which will completely revolutionize our knowledge of AGN and the origin of the CXB. 

Improved sensitivity could be achieved by increasing the effective area (through more optical modules and mirror shells), lowering the instrumental background, and perhaps most importantly, decreasing the size of the PSF to avoid source confusion and reduce background. Such a configuration would increase the number of sources by a factor of $\sim$70 (assuming a similar survey strategy used by {\it NuSTAR}) and allow us to probe a factor of 30 deeper in the 8--24 keV band, directly resolving as much as 80\% of the CXB, a regime that so far has only been probed in stacking studies (see Figs. \ref{lxz} and 3). {\em HEX-P} would also allow us to move to higher energies (to 50 keV and beyond), where no detection has been found so far by {\it NuSTAR} (see Masini et al. 2018). 

With increased high-energy bandwidth and  area (as proposed for {\em HEX-P}), a {\em NuSTAR}-like survey  (Fig.~3) would allow us to accurately measure $N_{\rm H}$ for even heavily obscured, distant AGN (up to $z=3$) and constrain the redshift evolution of the $N_{\rm H}$ distribution to extreme Compton-thick columns. Survey exposures  would yield $\sim$2,000 counts for AGN at $0.5<z<2$ and $L_X>10^{43.5}$ erg s$^{-1}$, constraining the high-$z$ cut off to 30\% error (Figure \ref{coronaT}). Building on the ground-breaking  analyses of {\it NuSTAR} with {\it Chandra} and {\it XMM-Newton}, {\em HEX-P} would provide excellent synergy with future soft X-ray missions (e.g., {\em XRISM}, {\em Athena}, {\em Lynx}). These joint observations would take particular advantage of the high spectral resolution of soft X-ray calorimeters, providing exquisite measurements of X-ray line complexes that, together with hard X-ray constraints on the Compton hump, are critical for understanding the geometry of obscuration and the physics of accretion.


\pagebreak
\noindent \textbf{References}\\

\noindent 
$\bullet$ {\color{black} Aalto, F., Battersby, C.  et al. 2019 {\it Astro2020 Science White Paper: Obscured galaxy nuclei -- hidden AGNs and extreme starbursts}}\\
$\bullet$ Aird, J., Alexander, D.~,M., Ballantyne, D.~R., et al. 2015a, ApJ, 815, 66 \\
$\bullet$ Aird, J., Coil, A.~L., Georgakakis, A., et al. 2015b, MNRAS, 451, 1892 \\
$\bullet$ 	
Ajello, M., Greiner, J., Sato, G., et al. 2008, ApJ, 689, 666\\
$\bullet$ Ananna, T., Treister, E., Urry, E.~M., et al.\ 2019, ApJ, 871, 240\\
$\bullet$ Balokovi\'{c}, M., Brightman, M., Harrison, F.~A., et al. 2018, ApJ, 854, 42\\
$\bullet$ Bauer, F.~E., Ar{\'e}valo, P., Walton, D.~J., et al.\ 2015, ApJ, 812, 116 \\
$\bullet$ Blecha, L., Snyder, G.~F., Satyapal, S., \& Ellison, S.~L. 2018, MNRAS 478, 3056\\
$\bullet$ Buchner, J. \& Bauer F.~E. 2017, MNRAS, 465,4348\\
$\bullet$ Civano, F., Capelluti, N., Hickox, R.~C. et al. 2019, {\it Astro2020 Science White Paper: Cosmic evolution of supermassive black holes: A view into the next two decades}\\
$\bullet$ Comastri A., Gilli R., Marconi A., et al., 2015. A\&A 574, L10\\
$\bullet$ Del Moro, A., Alexander, D.~M., Aird, J.~A., et al. 2017, ApJ, 849, 57\\
$\bullet$ Fabian, A.~C., Lohfink, A., Kara, E., et al.  2015, MNRAS, 451, 4375\\
$\bullet$ Gandhi, P., Lansbury, G.~B., Alexander, D.~M. et al.\ 2014, ApJ, 792, 117\\
$\bullet$ Giacconi, R., Gursky, H., Paolini, F.~R., \& Rossi, B.~B. 1962,  Phys. Rev. Lett. 9, 439\\
$\bullet$ Gilli, R., Comastri, A., \& Hasinger, G. 2007, A\&A, 463, 79\\
$\bullet$ Jones, M.~L., Hickox, R.~C., Mutch, S.~J., et al.\ 2017, ApJ, 843, 125 \\
$\bullet$ Harrison, F.~A., Craig, W.~W., Christensen, F.~E., et al.\ 2013, ApJ, 770, 103 \\ 
$\bullet$ Harrison, F.~A., Aird, J., Civano, F., et al.  2016, ApJ 831, 185\\
{\color{black}$\bullet$ 
Madsen, K.~K., Harrison, F., Broadway, D., et al., Proc. SPIE 10699, Space Telescopes and Instrumentation 2018: Ultraviolet to Gamma Ray, 106996M \\$\bullet$ Hickox, R.~C. \& Alexander, D.~M. 2018, ARA\&A, 56, 625\\
$\bullet$ Hickox, R.~C. \& Markevitch, M. 2006, ApJ, 645, 95\\
$\bullet$ Kamraj N. et al. 2019, {\it Astro2020 Science White Paper: Probing the Physical Properties of the Corona in Accreting Black Holes}\\
$\bullet$ Koss, M.~J., Assef, R., Balokovi\'{c}, et al. 2016, ApJ, 825, 85\\
$\bullet$ Lansbury, G.~B., Alexander, D.~M., Del Moro, A., et al. 2014, ApJ, 785, 17\\
$\bullet$ Lansbury, G.~B., Gandhi P., Alexander, D.~M., et al.  2015, ApJ, 805, 115\\
$\bullet$ Lansbury, G.~B., Alexander, D.~M., Aird, J., et al.\ 2017, ApJ, 846, 20\\
$\bullet$ Lanz, L., Hickox, R.~C., Balokovi\'{c}, M., et al. 2019, ApJ, 870, 26\\
$\bullet$ Lanzuisi, G., et al. 2019, ApjL, 875, L20 \\
$\bullet$ Luo, B., Brandt W.~N., Alexander D.~M., et al. 2014, ApJ, 794, 70\\
$\bullet$ Luo, B., Brandt W.~N., Xue, Y.~Q., et al. 2017, ApJS, 228, 2\\
$\bullet$ Marchesi, S., Ajello, M., Marcotulli, L., et al. 2018, ApJ, 854, 49\\
$\bullet$ Marchesi, S., Ajello, M., Zhao, X., et al.\ 2019, ApJ, 872, 8\\
$\bullet$ Masini, A., Comastri, A., Civano, F., et al. 2018, ApJ, 867, 162\\
$\bullet$ Matt, G, Balokovi\'{c}, M., Marinucci, A. et al.  2015, MNRAS, 447, 3029\\
$\bullet$ Netzer, H. 2015, ARA\&A, 53, 365\\
$\bullet$ Ricci, C., Bauer, F.~E., Treister, E., 2017a, MNRAS, 468, 1273\\
$\bullet$ Ricci, C., Trakhtenbrot, B., Koss, M.~J., et al. 2017b, Nature, 549, 488\\
$\bullet$ Ricci, C., Ho, L.~C., Fabian, A.~C., et al., 2018, MNRAS, 480, 1819\\
$\bullet$ Risaliti, G., Harrison, F.~A., Madsen, K.~K., et al.\ 2013, Nature, 494, 449 \\
$\bullet$ Stern, D., Lansbury, G.~B., Assef, R~J., et al. 2014, ApJ, 749, 102\\
$\bullet$ Teng, S.~H., Rigby, J.~R., Stern, D., et al.\ 2015, ApJ, 814, 56\\
$\bullet$ Ueda, Y., Akiyama, M., Hasinger, G., et al. 2014, ApJ, 768, 104\\
$\bullet$ Wilman, R.~J. \& Fabian, A.~C., MNRAS, 309, 862\\
$\bullet$ Xue, Y.~Q., Wang, S.~X., Brandt, W.N., et al. 2012, ApJ, 758, 129\\
$\bullet$ Yan, W., Hickox, R.~C., Hainline, K.~N., et al. 2019, ApJ, 870, 33\\
$\bullet$ Zappacosta, L., Comastri, A., Civano, F., et al.\ 2018, ApJ, 854, 33 

\end{document}